\documentclass[aps,prl,showpacs,twocolumn,lineno,groupedaddress]{revtex4}  
\usepackage{graphicx}  
\usepackage{dcolumn}   
\usepackage{bm}        
\usepackage{amssymb}   

\newcommand{\mt}{\mbox{${m}_{T}$}}

\newcommand{\mll}{\mbox{${m}_{\ell\ell}$}}

\newcommand{\mte}{\mbox{${m}_{T}^{e}$}}
\newcommand{\mtel}{\mbox{${m}_{T}^{e_{1}}$}}
\newcommand{\mtenl}{\mbox{${m}_{T}^{e_{2}}$}}
\newcommand{\mtm}{\mbox{${m}_{T}^{\mu}$}}

\newcommand{\pt}{\mbox{$p_{T}$}}

\newcommand{\pte}{\mbox{$p_{T}^{e}$}}

\newcommand{\Etj}{\mbox{$E_{T}^{\text{jet}}$}}

\newcommand{\lint}{\mbox{$\int{{\cal L} \mathrm{d}t}$}}

\newcommand{\ipb}{\mbox{$\text{pb}^{-1}$}}

\newcommand{\Eslash}{\mbox{$E \kern-0.6em\slash$}}
\newcommand{\etmiss}{\mbox{$\Eslash_T$}}

\newcommand{\pslash}{\mbox{$p \kern-0.5em\slash$}}

\newcommand{\etmisspar}[1]{\mbox{$\Eslash_T^{\text{#1}}$}}

\newcommand{\dphimm}{\mbox{$\Delta \phi_{\mu\mu}$}}

\newcommand{\zg}{\mbox{$Z/\gamma^*$}}
\newcommand{\zgee}{\mbox{$Z/\gamma^*\rightarrow ee$}}

\newcommand{\zgmm}{\mbox{$Z/\gamma^*\rightarrow\mu\mu$}}
\newcommand{\zgtt}{\mbox{$Z/\gamma^*\rightarrow\tau\tau$}}

\newcommand{\ttbar}{\mbox{$t\bar{t}$}}
\newcommand{\ppbar}{\mbox{$p\bar{p}$}}

\newcommand{\ee}{\mbox{$e^{+}e^{-}$}}
\newcommand{\emu}{\mbox{$e^{\pm}\mu^{\mp}$}}
\newcommand{\mumu}{\mbox{$\mu^{+}\mu^{-}$}}

\newcommand{\GeV}{\text{GeV}}

\begin{document}

\hspace{5.2in} \mbox{Fermilab-Pub-04/293-E}

\title{Measurement of the $WW$ production cross section in \ppbar\ collisions
  at $\sqrt{s} = 1.96$ TeV}
%
\author{                                                                      
V.M.~Abazov,$^{33}$                                                           
B.~Abbott,$^{70}$                                                             
M.~Abolins,$^{61}$                                                            
B.S.~Acharya,$^{27}$                                                          
M.~Adams,$^{48}$                                                              
T.~Adams,$^{46}$                                                              
M.~Agelou,$^{17}$                                                             
J.-L.~Agram,$^{18}$                                                           
S.H.~Ahn,$^{29}$                                                              
M.~Ahsan,$^{55}$
% ...
\\                                                                            
\vskip 0.30cm                                                                 
\centerline{(D\O\ Collaboration)}                                             
\vskip 0.30cm                                                                 
}                                                                             
\address{                                                                     
\centerline{$^{1}$Universidad de Buenos Aires, Buenos Aires, Argentina}       
\centerline{$^{2}$LAFEX, Centro Brasileiro de Pesquisas F{\'\i}sicas,         
                  Rio de Janeiro, Brazil}                                     
\centerline{$^{3}$Universidade do Estado do Rio de Janeiro,                   
                  Rio de Janeiro, Brazil}                                     
\centerline{$^{4}$Instituto de F\'{\i}sica Te\'orica, Universidade            
                  Estadual Paulista, S\~ao Paulo, Brazil}                     
\centerline{$^{5}$Simon Fraser University, Burnaby, Canada, University of     
                  Alberta, Edmonton, Canada,}                                 
\centerline{McGill University, Montreal, Canada and York University,          
                  Toronto, Canada}                                            
\centerline{$^{6}$Institute of High Energy Physics, Beijing,                  
                  People's Republic of China}                                 
\centerline{$^{7}$Universidad de los Andes, Bogot\'{a}, Colombia}             
\centerline{$^{8}$Charles University, Center for Particle Physics,            
                  Prague, Czech Republic}                                     
\centerline{$^{9}$Czech Technical University, Prague, Czech Republic}         
\centerline{$^{10}$Institute of Physics, Academy of Sciences, Center          
                  for Particle Physics, Prague, Czech Republic}               
\centerline{$^{11}$Universidad San Francisco de Quito, Quito, Ecuador}        
\centerline{$^{12}$Laboratoire de Physique Corpusculaire, IN2P3-CNRS,         
                 Universit\'e Blaise Pascal, Clermont-Ferrand, France}        
\centerline{$^{13}$Laboratoire de Physique Subatomique et de Cosmologie,      
                  IN2P3-CNRS, Universite de Grenoble 1, Grenoble, France}     
\centerline{$^{14}$CPPM, IN2P3-CNRS, Universit\'e de la M\'editerran\'ee,     
                  Marseille, France}                                          
\centerline{$^{15}$Laboratoire de l'Acc\'el\'erateur Lin\'eaire,              
                  IN2P3-CNRS, Orsay, France}                                  
\centerline{$^{16}$LPNHE, Universit\'es Paris VI and VII, IN2P3-CNRS,         
                  Paris, France}                                              
\centerline{$^{17}$DAPNIA/Service de Physique des Particules, CEA, Saclay,    
                  France}                                                     
\centerline{$^{18}$IReS, IN2P3-CNRS, Universit\'e Louis Pasteur, Strasbourg,  
                  France and Universit\'e de Haute Alsace, Mulhouse, France}  
\centerline{$^{19}$Institut de Physique Nucl\'eaire de Lyon, IN2P3-CNRS,      
                   Universit\'e Claude Bernard, Villeurbanne, France}         
\centerline{$^{20}$RWTH Aachen, III. Physikalisches Institut A,               
                   Aachen, Germany}                                           
\centerline{$^{21}$Universit{\"a}t Bonn, Physikalisches Institut,             
                  Bonn, Germany}                                              
\centerline{$^{22}$Universit{\"a}t Freiburg, Physikalisches Institut,         
                  Freiburg, Germany}                                          
\centerline{$^{23}$Universit{\"a}t Mainz, Institut f{\"u}r Physik,            
                  Mainz, Germany}                                             
\centerline{$^{24}$Ludwig-Maximilians-Universit{\"a}t M{\"u}nchen,            
                   M{\"u}nchen, Germany}                                      
\centerline{$^{25}$Fachbereich Physik, University of Wuppertal,               
                   Wuppertal, Germany}                                        
\centerline{$^{26}$Panjab University, Chandigarh, India}                      
\centerline{$^{27}$Tata Institute of Fundamental Research, Mumbai, India}     
\centerline{$^{28}$University College Dublin, Dublin, Ireland}                
\centerline{$^{29}$Korea Detector Laboratory, Korea University,               
                   Seoul, Korea}                                              
\centerline{$^{30}$CINVESTAV, Mexico City, Mexico}                            
\centerline{$^{31}$FOM-Institute NIKHEF and University of                     
                  Amsterdam/NIKHEF, Amsterdam, The Netherlands}               
\centerline{$^{32}$University of Nijmegen/NIKHEF, Nijmegen, The               
                  Netherlands}                                                
\centerline{$^{33}$Joint Institute for Nuclear Research, Dubna, Russia}       
\centerline{$^{34}$Institute for Theoretical and Experimental Physics,        
                  Moscow, Russia}                                             
\centerline{$^{35}$Moscow State University, Moscow, Russia}                   
\centerline{$^{36}$Institute for High Energy Physics, Protvino, Russia}       
\centerline{$^{37}$Petersburg Nuclear Physics Institute,                      
                   St. Petersburg, Russia}                                    
\centerline{$^{38}$Lund University, Lund, Sweden, Royal Institute of          
                   Technology and Stockholm University, Stockholm,            
                   Sweden and}                                                
\centerline{Uppsala University, Uppsala, Sweden}                              
\centerline{$^{39}$Lancaster University, Lancaster, United Kingdom}           
\centerline{$^{40}$Imperial College, London, United Kingdom}                  
\centerline{$^{41}$University of Manchester, Manchester, United Kingdom}      
\centerline{$^{42}$University of Arizona, Tucson, Arizona 85721, USA}         
\centerline{$^{43}$Lawrence Berkeley National Laboratory and University of    
                  California, Berkeley, California 94720, USA}                
\centerline{$^{44}$California State University, Fresno, California 93740, USA}
\centerline{$^{45}$University of California, Riverside, California 92521, USA}
\centerline{$^{46}$Florida State University, Tallahassee, Florida 32306, USA} 
\centerline{$^{47}$Fermi National Accelerator Laboratory, Batavia,            
                   Illinois 60510, USA}                                       
\centerline{$^{48}$University of Illinois at Chicago, Chicago,                
                   Illinois 60607, USA}                                       
\centerline{$^{49}$Northern Illinois University, DeKalb, Illinois 60115, USA} 
\centerline{$^{50}$Northwestern University, Evanston, Illinois 60208, USA}    
\centerline{$^{51}$Indiana University, Bloomington, Indiana 47405, USA}       
\centerline{$^{52}$University of Notre Dame, Notre Dame, Indiana 46556, USA}  
\centerline{$^{53}$Iowa State University, Ames, Iowa 50011, USA}              
\centerline{$^{54}$University of Kansas, Lawrence, Kansas 66045, USA}         
\centerline{$^{55}$Kansas State University, Manhattan, Kansas 66506, USA}     
\centerline{$^{56}$Louisiana Tech University, Ruston, Louisiana 71272, USA}   
\centerline{$^{57}$University of Maryland, College Park, Maryland 20742, USA} 
\centerline{$^{58}$Boston University, Boston, Massachusetts 02215, USA}       
\centerline{$^{59}$Northeastern University, Boston, Massachusetts 02115, USA} 
\centerline{$^{60}$University of Michigan, Ann Arbor, Michigan 48109, USA}    
\centerline{$^{61}$Michigan State University, East Lansing, Michigan 48824,   
                   USA}                                                       
\centerline{$^{62}$University of Mississippi, University, Mississippi 38677,  
                   USA}                                                       
\centerline{$^{63}$University of Nebraska, Lincoln, Nebraska 68588, USA}      
\centerline{$^{64}$Princeton University, Princeton, New Jersey 08544, USA}    
\centerline{$^{65}$Columbia University, New York, New York 10027, USA}        
\centerline{$^{66}$University of Rochester, Rochester, New York 14627, USA}   
\centerline{$^{67}$State University of New York, Stony Brook,                 
                   New York 11794, USA}                                       
\centerline{$^{68}$Brookhaven National Laboratory, Upton, New York 11973, USA}
\centerline{$^{69}$Langston University, Langston, Oklahoma 73050, USA}        
\centerline{$^{70}$University of Oklahoma, Norman, Oklahoma 73019, USA}       
\centerline{$^{71}$Brown University, Providence, Rhode Island 02912, USA}     
\centerline{$^{72}$University of Texas, Arlington, Texas 76019, USA}          
\centerline{$^{73}$Southern Methodist University, Dallas, Texas 75275, USA}   
\centerline{$^{74}$Rice University, Houston, Texas 77005, USA}                
\centerline{$^{75}$University of Virginia, Charlottesville, Virginia 22901,   
                   USA}                                                       
\centerline{$^{76}$University of Washington, Seattle, Washington 98195, USA}  
}                                                                             
\date{\today}

\begin{abstract}
We present a measurement of the $W$ boson pair-production cross section in 
\ppbar\ collisions at a center-of-mass energy of $\sqrt{s}=1.96$ TeV. 
The data, collected with the Run II D\O\ detector, correspond 
to an integrated luminosity of 224--252 \ipb\ depending on the final state 
(\ee, \emu\ or \mumu).
We observe 25 candidates with a 
background expectation of $8.1\pm0.6(\rm stat) \pm 0.6 (\rm syst) \pm 0.5(\rm lum) $ events. 
The probability for an upward fluctuation of the background to produce the
observed signal is $2.3\times 10^{-7}$, equivalent to 5.2 standard deviations.
The measurement yields a cross section of
$13.8^{+4.3}_{-3.8}(\rm stat)^{~+1.2}_{~-0.9}(\rm syst) \pm 0.9(\rm lum)\ {\rm pb}$, in agreement with predictions from the standard model.
\end{abstract}

\maketitle


The measurement of the $W$ boson pair-production cross section 
$\sigma_{p\bar{p}\rightarrow W^+W^-}$\ offers a good opportunity to test
the non-Abelian structure of the standard model (SM). Furthermore, this measurement is sensitive
to new phenomena since anomalous trilinear couplings~\cite{couplings} or the
production and decay of new particles such as the
Higgs boson~\cite{higgs}
would enhance the rate of $W$ boson pair-production. 
The next-to-leading order (NLO) calculations for 
$\sigma_{p\bar{p}\rightarrow W^+W^-}$ \cite{ellis} predict a cross section
of 12.0--13.5 pb at $\sqrt{s}=1.96$ TeV. 
The CDF Collaboration reported evidence for $W$ boson pair-production, 
based on 108 \ipb\ of data collected in Run~I of the 
Fermilab Tevatron Collider at $\sqrt{s}=1.8$ TeV, 
with a cross section $\sigma_{p\bar{p}\to  W^+W^-} = 10.2^{+6.3}_{-5.1}(\rm stat) \pm 1.6(\rm syst)\ {\rm pb}$~\cite{wwcdf}.

In this Letter we present a measurement of the $W^+W^-$ production cross
section in leptonic final states 
$\ppbar \rightarrow W^+W^-\to \ell^+\nu\ell^-\bar{\nu}$ ($\ell= e, \mu$). 
We use data collected between April 2002 and
March 2004 in \ppbar\ collisions at $\sqrt{s} = 1.96$ TeV of Run II of the Tevatron 
Collider. The integrated luminosities 
are $252\pm 16$ \ipb, $235\pm 15$ \ipb, and $224\pm 15$ \ipb\ 
for the \ee, \emu, and \mumu\ channels, respectively. 
The differences in the integrated luminosities for various channels are
primarily due to different trigger conditions.

We briefly describe the main components of the D\O\ Run II 
detector \cite{d0det} important to this analysis. The central tracking system 
consists of a silicon microstrip tracker (SMT) and
a central fiber tracker (CFT), both located within a 2.0~T axial magnetic
field. The SMT strips have a typical pitch of 50--80 $\mu$m, and the design 
is optimized for tracking and vertexing over the pseudorapidity range 
$|\eta| < 3$, where $\eta = -\ln{(\tan{\theta \over 2})}$ with polar angle 
$\theta$. The system
has a six-barrel longitudinal structure, each with a set of four silicon layers
arranged axially around the beam pipe, interspersed with sixteen radial disks.
The CFT has eight thin coaxial barrels, each supporting two doublets of
overlapping scintillating fibers of 0.835~mm diameter, one doublet 
parallel to the beam axis, the other alternating by $\pm 3^{\circ}$ relative
to the beam axis.

A liquid-argon/uranium calorimeter surrounds the central tracking system and
consists of a central calorimeter (CC)
covering to $|\eta|$ $\approx 1.1$, and two end calorimeters (EC) 
extending coverage for $|\eta| < 4.2$, all housed in separate cryostats
\cite{d0cal}. Scintillators between the CC and EC cryostats provide sampling
of showers for 1.1 $< |\eta| <$ 1.4.

The muon system is located outside the calorimeters and consists of a layer
of tracking detectors and scintillation trigger counters inside toroid magnets
which provide a 1.8~T magnetic field, followed by two similar layers 
behind each toroid. Tracking in
the muon system for $|\eta| < 1$ relies on 10~cm wide drift tubes \cite{d0cal},
while 1~cm mini-drift tubes are used for $1 < |\eta| < 2$
~\cite{Abramov:1998ti}.

The $W^+W^-\to \ell^+\nu\ell^-\bar{\nu}$ candidates are selected by triggering 
on single or di-lepton events using a three level trigger system. 
The first trigger
level uses hardware to select electron candidates based on energy deposition in
the electromagnetic part of the calorimeter and selects muon candidates formed
by hits in two layers of the muon scintillator system. Digital signal
processors in the second trigger level form muon track candidate segments 
defined
by hits in the muon drift chambers and scintillators. At the third level,
software algorithms running on a computing farm and exploiting the full event
information are used to make the final selection of events which are recorded
for offline analysis. 

In the further offline analysis electrons are identified by electromagnetic 
showers in the calorimeter.
These showers are chosen by comparing the longitudinal and
transverse shower profiles to those of simulated electrons.
The showers must be isolated, deposit most of their energy in the 
electromagnetic part of  the calorimeter, and pass a likelihood
criterion that includes a spatial track match and,
in the CC region, an $E/p$ requirement, where $E$ is the energy of the
calorimeter cluster and $p$ is the
momentum of the track. All electrons are required to be in the
pseudorapidity range $|\eta| < 3.0$. 
The transverse momentum measurement of the electrons is based on calorimeter 
cell energy information.

Muon tracks are reconstructed from 
hits in the wire chambers and scintillators in the muon system and must 
match a track in the central tracker. 
To select isolated muons, the scalar sum of the transverse momentum of
all tracks other than that of the muon in a cone of ${\cal R} = 0.5$ around the
muon track must be less than 4 GeV, where ${\cal R} =
\sqrt{(\Delta\phi)^2+(\Delta\eta)^2}$ and $\phi$ is the azimuthal angle.
Muon detection is restricted to the coverage of the muon system 
$|\eta| < 2.0$.
Muons from cosmic rays are rejected by requiring a timing 
criterion on the hits in the scintillator layers 
as well as applying restrictions on the position of the
muon track with respect to the primary vertex.  
 
The decay of two $W$ bosons into electrons or muons results in three
different final states $\ee+X$ ($ee$\ channel), $\emu+X$ ($e\mu$\ channel), and
$\mumu+X$ ($\mu\mu$\ channel), each of which consists of two oppositely
charged isolated high transverse momentum, $p_T$, leptons and large missing 
transverse energy, \etmiss , due to the escaping neutrinos.
The selection criteria for each channel were chosen to maximize the
the expected signal significance, while keeping high efficiency for
$WW$ production.

\begin{figure}[t]
\includegraphics[width=0.49\columnwidth]{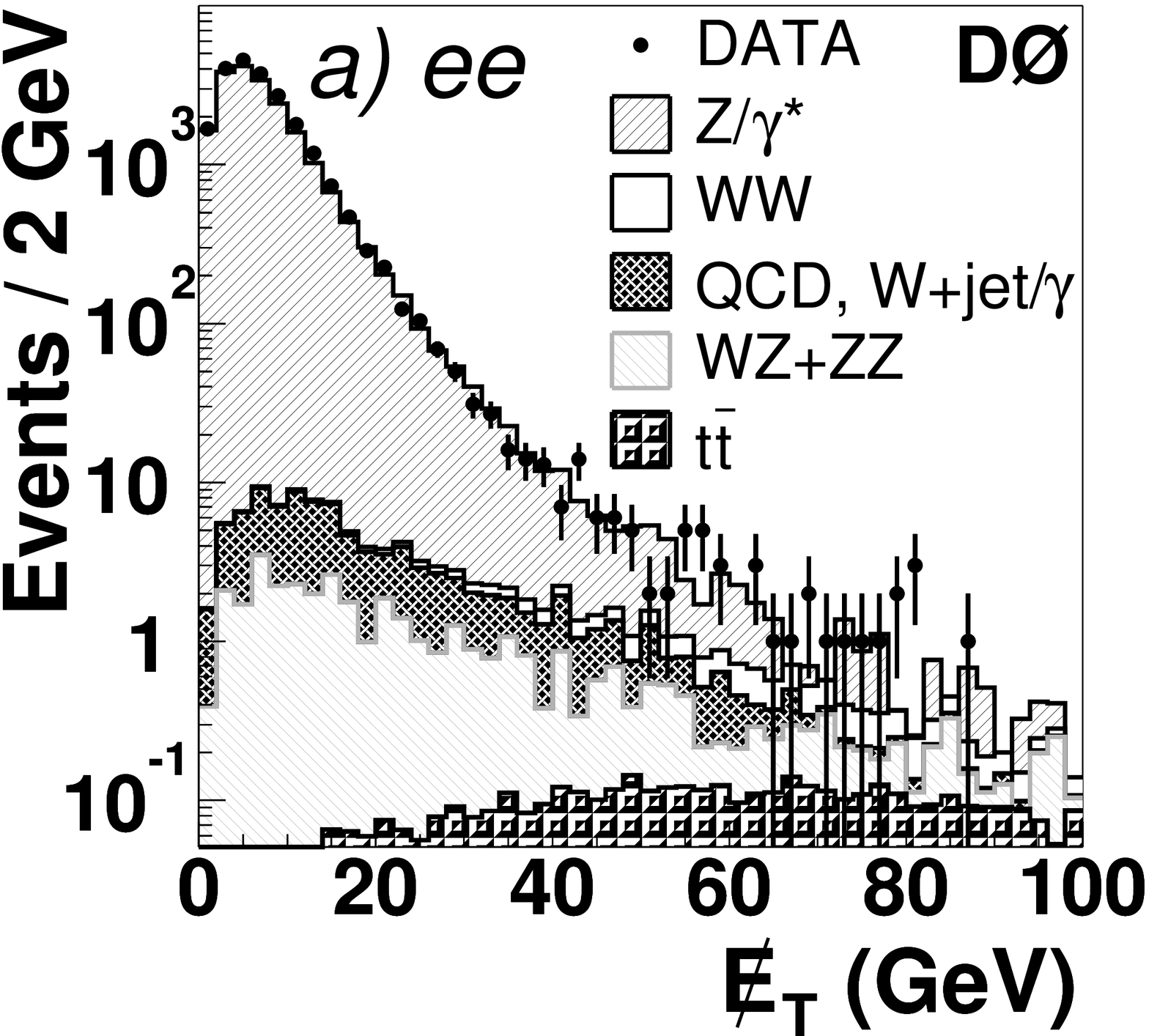}\hfill
\includegraphics[width=0.49\columnwidth]{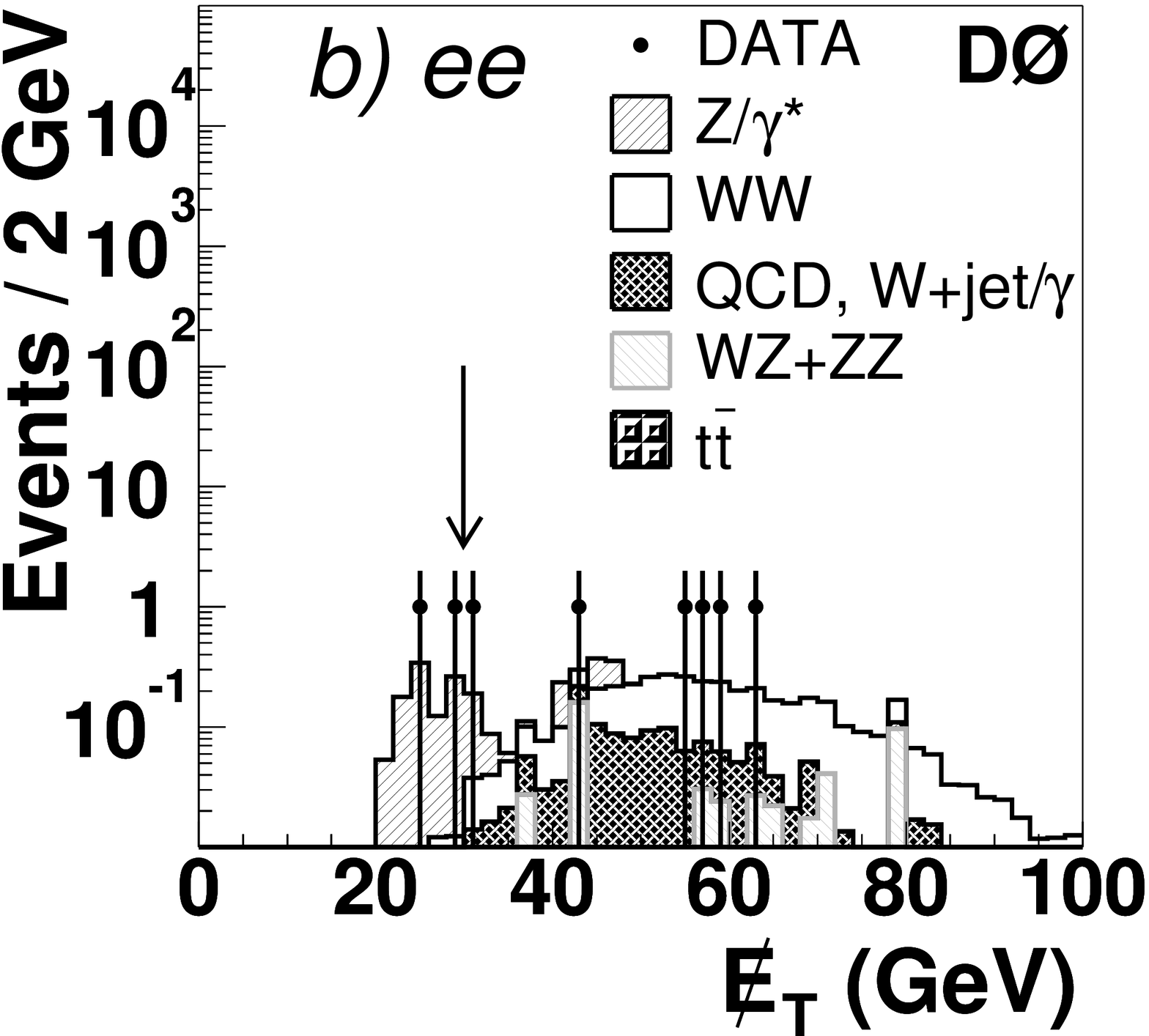}
\includegraphics[width=0.49\columnwidth]{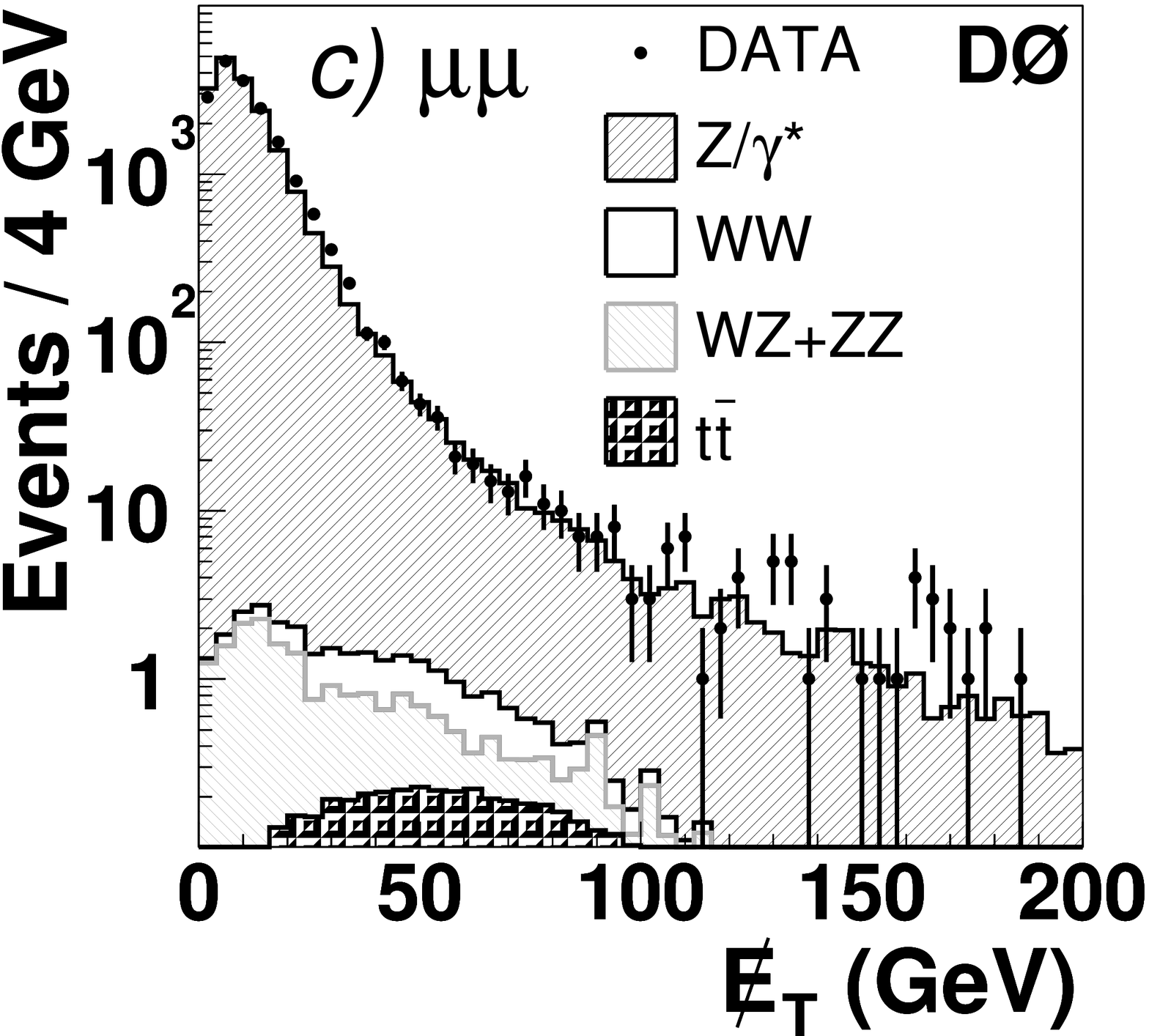}\hfill
\includegraphics[width=0.49\columnwidth]{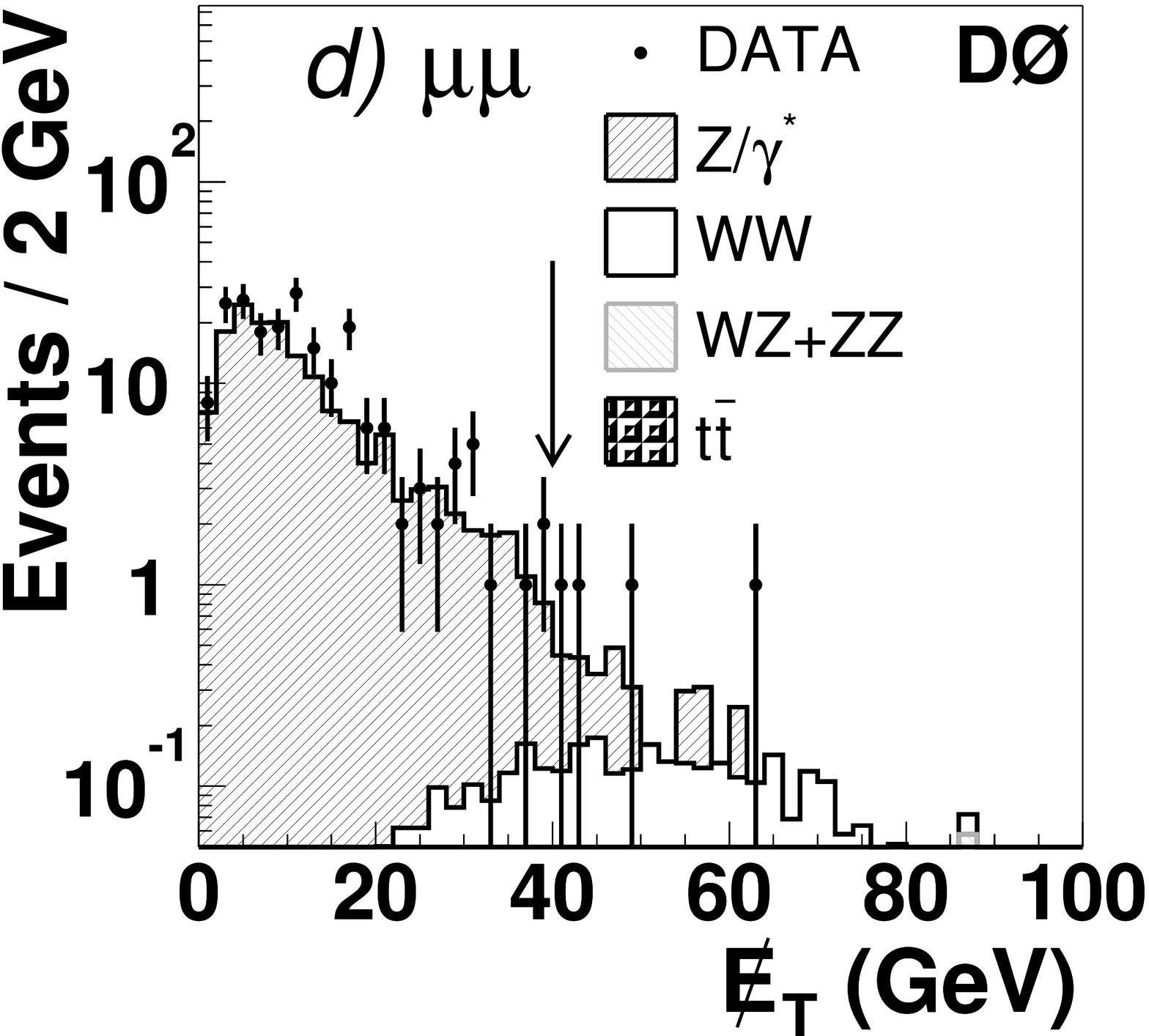}
\includegraphics[width=0.49\columnwidth]{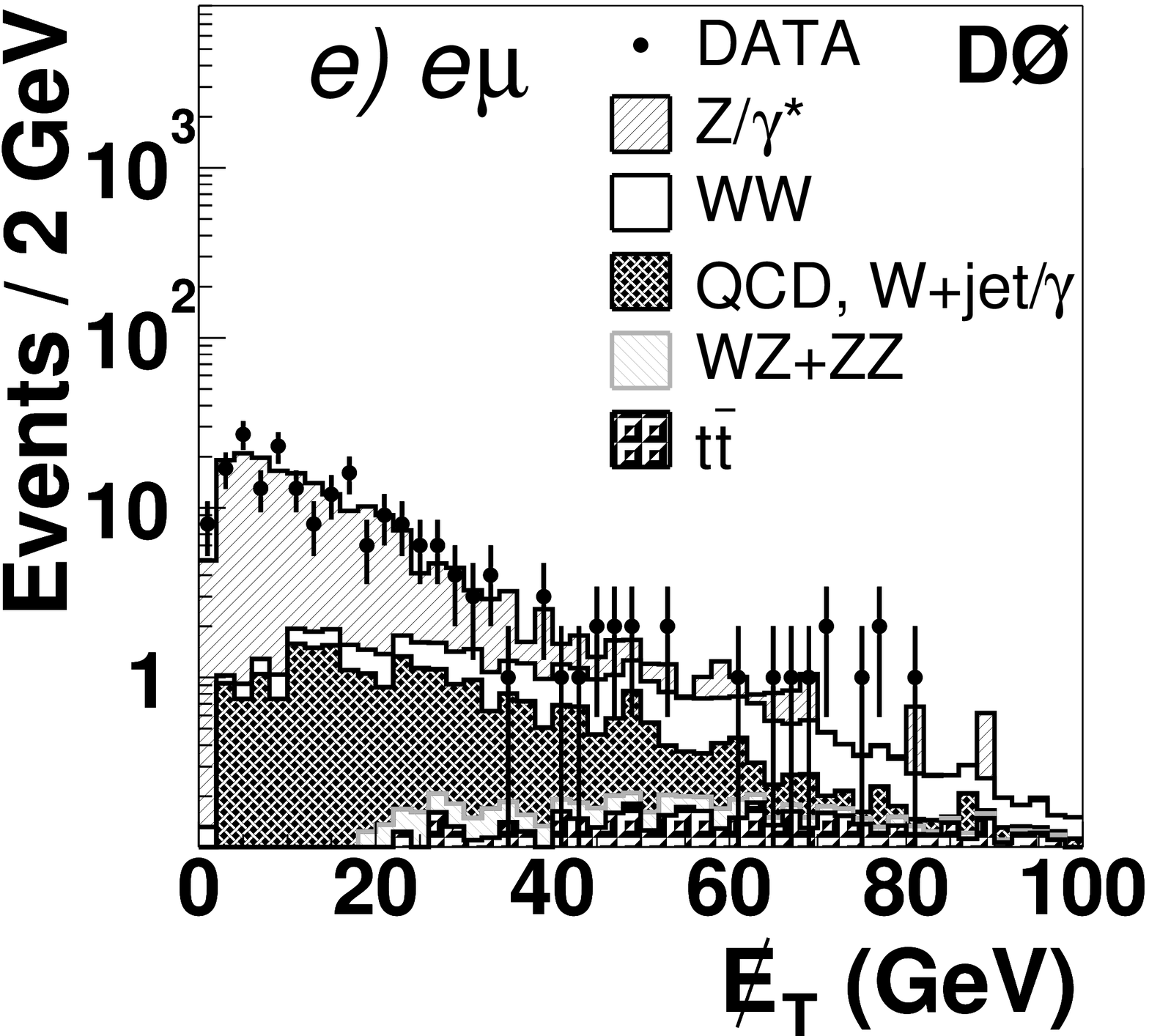}\hfill
\includegraphics[width=0.49\columnwidth]{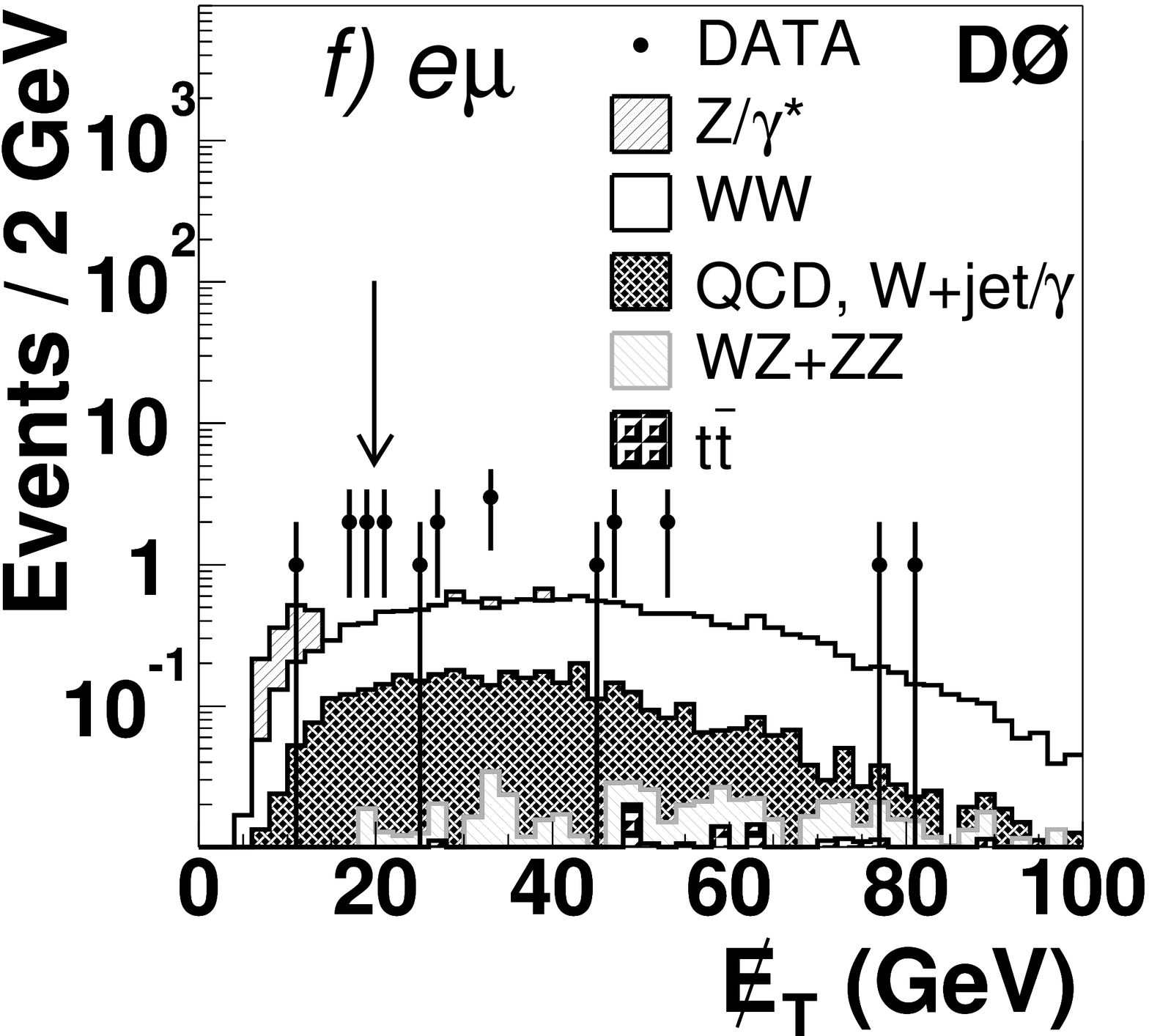}
\caption{\label{fig:met_plots} Distribution of the missing transverse 
  energy \etmiss\ after applying the initial transverse 
  momentum cuts in the (a) $ee$, (c) $\mu\mu$, and (e) $e\mu$ channel.
  Figures (b), (d), and (f) show the \etmiss\ distributions after the final 
  selection except for the 
  \etmiss\ criterion for the $ee$, $\mu\mu$, and $e\mu$ channel, respectively.
  The arrows indicate the cut values.
  QCD contribution is negligible in Figs. (c) and (d). }
\end{figure}

In all three channels, two leptons originating from the same vertex
are required to be of opposite charge, and must have
\pt\ $>$ 20~\GeV\ for the leading lepton and \pt\ $>$ 15~\GeV\ for 
the trailing one.
Figure~\ref{fig:met_plots} shows the good agreement between data and
Monte Carlo (MC) in \etmiss\ distributions for 
the $ee$\ channel (a), the $\mu\mu$\ channel (c) and the $e\mu$ channel (e) 
after applying the lepton transverse momentum cuts.
In all cases, the background is largely dominated by $Z/\gamma^*$ production
which is suppressed by requiring the \etmiss\ to be greater than 30~GeV, 
40~GeV, and 20~GeV in the $ee$, $\mu\mu$, and $e\mu$\ channels, respectively. 
The different cut values among the three channels are due to the different
momentum resolution of electrons and muons.

\begin{figure}[t]
\includegraphics[width=0.49\columnwidth]{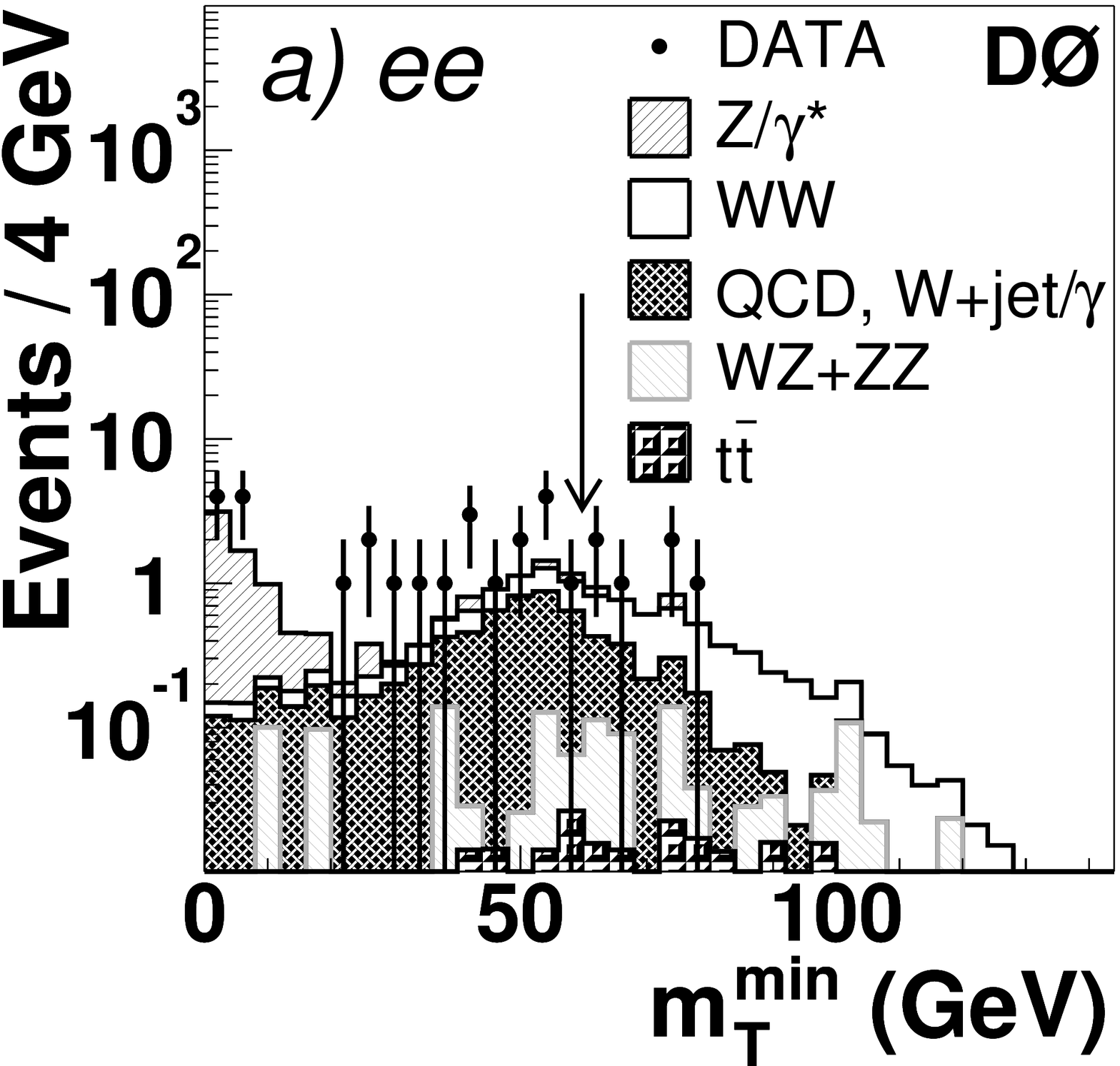}\hfill
\includegraphics[width=0.49\columnwidth]{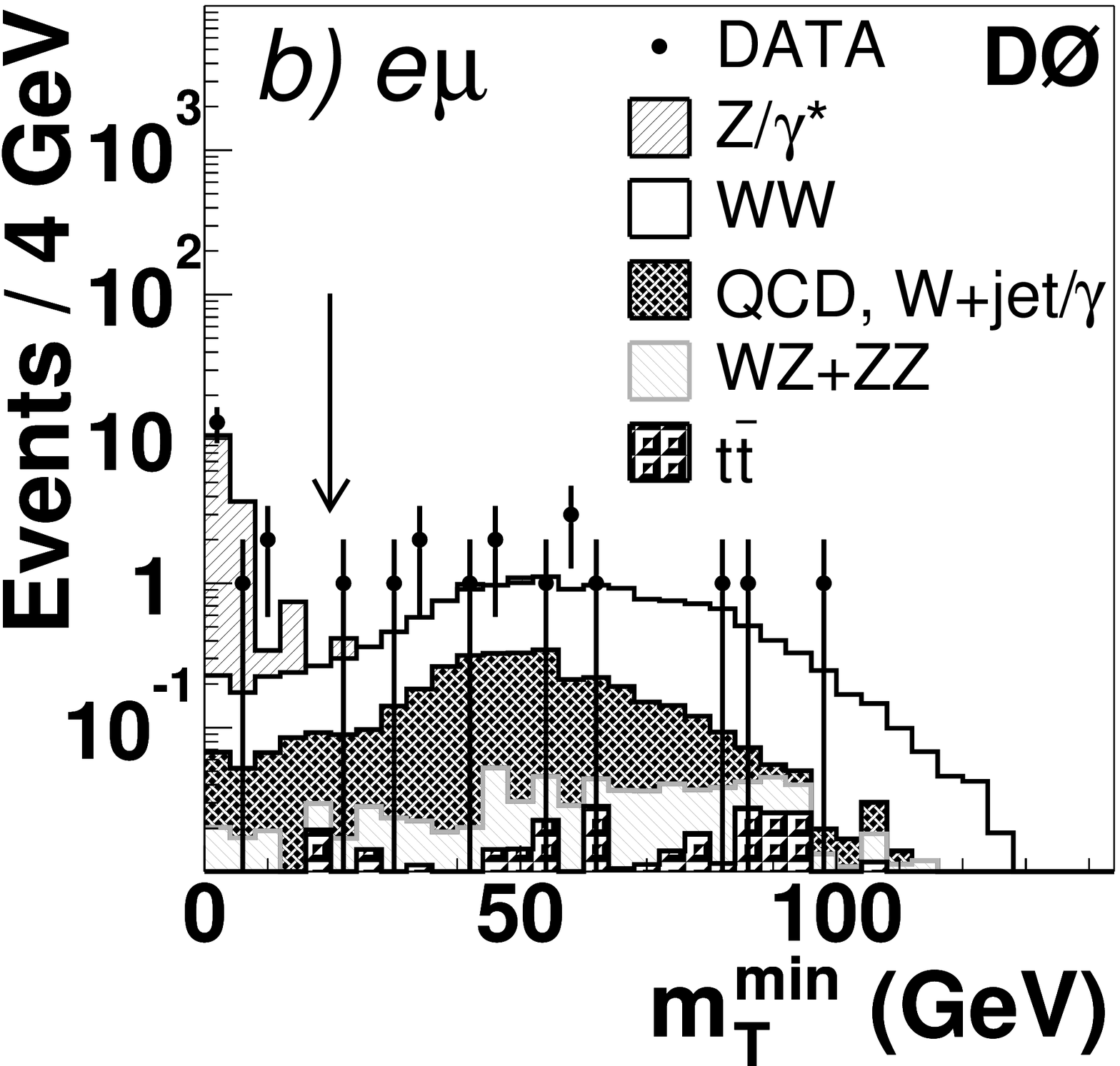}
\caption{\label{fig:mt_plots} Distribution of the minimal transverse mass
         $m_T^{\rm min}$ after applying all selection criteria except the cut 
         on $m_T^{\rm min}$  for the (a) $ee$ and the (b) $e\mu$ channel.
     The arrows indicate the cut values.}
\end{figure}

In the $ee$ channel, additional cuts are applied to further reduce the 
$Z/\gamma^*$ background and other backgrounds. The minimal transverse mass 
$m_T^{\text{min}}=\min(\mtel, \mtenl)$ must exceed 60~\GeV, where 
\mt\ = $\sqrt{2\etmiss\pte(1-\cos\Delta\phi(\pte,\etmiss))}$.
Events are removed if the invariant di-electron mass is between 76 and
106~\GeV. Events are also removed if the \etmiss\ has a large contribution 
from the mismeasurement of jet energy, using the following procedure.
The fluctuation in the measurement of jet energy in the transverse plane 
can be approximated by $\Delta E^{\rm jet}\cdot\sin\theta^{\rm jet}$ where
$\Delta E^{\rm jet}$ is proportional to $\sqrt{{E}^{\rm jet}}$. 
The opening angle $\Delta\phi\left({\rm jet},\etmiss\right)$\ between this 
projected energy fluctuation and the missing transverse energy
provides a measure of the contribution of the jet to the missing transverse 
energy. The scaled missing transverse energy defined as
\begin{equation}
\etmisspar{Sc} =
\frac{\etmiss}{\sqrt{\sum_{\rm jets}\left(\Delta E^{\rm jet}\cdot\sin\theta^{\rm jet}\cdot\cos\Delta\phi\left({\rm jet},\etmiss\right)\right)^{2}}}
\end{equation}
is required to be greater than 15. Finally, to suppress 
the background from \ttbar\ production, the scalar sum of the transverse 
energies of all jets with \Etj\ $>$ 20~\GeV\ and $|\eta| < 2.5$, $H_T$,  is 
required to be less than 50~\GeV. 
Figure~\ref{fig:met_plots}(b) shows the \etmiss\ distribution 
after the final selection without applying the 
\etmiss\ criterion for the $ee$\ channel, and
Fig.~\ref{fig:mt_plots}(a) shows the distribution of the 
minimal transverse mass after applying all selection criteria except the cut 
on the minimal transverse mass. 
Six events remain in the $ee$ data sample after all of these cuts are 
applied.

For the $\mu\mu$ channel, to further reduce the \zg\ background, 
only events with an invariant di-muon mass between 
20 and 80~\GeV\ are retained.
Since the momentum resolution is worsening for high \pt\ tracks, 
an additional constrained fit is performed to reject events  
compatible with $Z$ boson production. 
The opening angle between the two muons in the transverse plane is 
required to be \dphimm\ $<$ 2.4. 
Finally, requiring $H_T <$ 100~\GeV\ removes the remaining background from 
\ttbar\ events. 
Figure~\ref{fig:met_plots}(d) shows the \etmiss\ distribution 
after the final selection without applying the 
\etmiss\ criterion for the $\mu\mu$\ channel.
Four events are observed in the $\mu\mu$ data sample after application of all 
selection criteria. 

In the $e\mu$ channel, to suppress the $WZ$ and $ZZ$ background, 
events are rejected if a third lepton is found and the invariant mass of 
two leptons of the same flavor and opposite charge is in the range from 
61 to 121~\GeV. To remove background 
from multijet production and \zgtt\ events, 
the minimal transverse mass $m_T^{\text{min}}=\min(\mte, \mtm)$ must exceed 
20~\GeV. The remaining \zgtt\ events are suppressed by removing events with
\etmisspar{Sc} $<$ 15.
Requiring $H_T <$ 50~\GeV\ rejects most of the \ttbar\ events.
To remove $W+\gamma$ events 
in which photons convert to electron-positron pairs, at least three hits in the
 silicon tracker are required for the electron track if the transverse mass 
determined from the muon and \etmiss\ is consistent
with the $W$ boson transverse mass. 
Figure~\ref{fig:met_plots}(f) shows the \etmiss\ distribution 
after the final selection without applying the 
\etmiss\ criterion for the $e\mu$\ channel, whereas 
Fig.~\ref{fig:mt_plots}(b) shows the distribution of the 
minimal transverse mass after applying all selection criteria except the cut 
on the minimal transverse mass. 
Fifteen events survive the final selection criteria in the $e\mu$ data sample.

The efficiency for $WW$ signal events to pass the acceptance and kinematic
criteria is determined using the {\sc pythia} 6.2 \cite{pythia} event 
generator followed by a detailed {\sc geant}-based \cite{geant}
simulation of the D\O\ detector. All trigger and reconstruction 
efficiencies are derived from the data. For the $ee$ channel, the
overall detection efficiency is  
($8.76\pm 0.13$)\%. The overall efficiencies for the $\mu\mu$ and $e\mu$ 
channels are ($6.22\pm 0.15$)\% and ($15.40\pm 0.20$)\%, respectively. Using an
NLO cross section of
13.5~pb \cite{ellis} and branching fractions $B$ of $0.1072\pm 0.0016$ for
$W\to e\nu$ and $0.1057\pm 0.0022$ for $W\to \mu\nu$ \cite{pdg}, the expected
number of events for the pair production of
$W$ bosons combined for all three channels is
$16.6\pm0.1(\rm stat) \pm 0.6(\rm syst) \pm 1.1(\rm lum) $ events, 
where the statistical error is given by the statistics of the MC sample.
The signal breakdown for the three channels is given by the
first line of Table~\ref{tab:events}.

Background contributions from \zg, $W$+jet/$\gamma$, \ttbar, $WZ$ and $ZZ$
events are estimated using the {\sc pythia} event generator. In addition, 
$W$+jet/$\gamma$ contributions are verified using {\sc alpgen} \cite{alpgen}.
All events are processed through the full detector simulation.
The background due to multijet production, when a jet is misidentified as an
electron, is determined from the data using a sample of like-sign
di-lepton events with inverted lepton quality cuts (called QCD background
in Figs.~\ref{fig:met_plots} and \ref{fig:mt_plots}).

For the normalization of \zg\ and
$W$+jet/$\gamma$ events, the NNLO cross sections from Ref.~\cite{hamberg} are 
used.
The cross section times branching ratio of \zg\ production 
in the invariant mass region 60\,\GeV $<$ \mll $<$ 130\,\GeV\ is 
$\sigma\times B = 254\ {\rm pb}$. For inclusive $W$ boson production with decays into 
a single lepton flavor state, this value is $\sigma\times B = 2717\ {\rm pb}$.
The NLO $WZ$ and $ZZ$  production cross section values are taken 
from Ref.~\cite{ellis} with $\sigma\times B = 0.014\ {\rm pb}$ for $WZ$ and
 $\sigma\times B = 0.002\ {\rm pb}$ for $ZZ$ production with decay into a 
single lepton flavor state. 
The calculations of Ref.~\cite{kidonakis} are used for \ttbar\ production
with $\sigma\times B = 0.076\ {\rm pb}$ with single flavor lepton decays of 
both $W$ bosons.
A summary of the background contributions together with signal
expectations and events observed in the data after the final
selection for the individual channels is shown in Table~\ref{tab:events}.
The total background sum is $8.1\pm0.6(\rm stat) \pm 0.6 (\rm syst)
\pm 0.5(\rm lum) $ events. The $e\mu$ channel has both
the highest signal efficiency and best signal-to-background ratio.
There is good agreement between the number of events
observed in the data and the sum of the expectations from $WW$ production and
the various backgrounds in all three channels.

\begin{table}[t]
\caption{\label{tab:events} Number of signal and background events expected
  and number of events observed after all
  selections are applied for the three channels. 
  Only statistical uncertainties are given.}
\begin{ruledtabular}
\begin{tabular}{lccc}
Process     & $ee$           & $e\mu$         & $\mu\mu$\\
\hline
$WW$ signal    & $3.42\pm 0.05$ & $11.10\pm 0.10$  & $2.10\pm 0.05$\\
\hline
\zgee          & $0.20\pm 0.06$ & ---            & ---\\
\zgmm          & ---            & $0.28\pm 0.09$ & $1.60\pm 0.40$\\
\zgtt          & $<0.01$     & $0.0\pm0.1$     & $<0.01$\\
\ttbar         & $0.18\pm 0.02$ & $0.34\pm 0.03$ & $0.09\pm 0.01$\\
$WZ$           & $0.33\pm 0.17$ & $0.38\pm 0.02$ & $0.15\pm 0.08$\\
$ZZ$           & $0.19\pm 0.06$ & $0.02\pm 0.02$ & $0.10\pm 0.04$\\
$W$+jet/$\gamma$ & $1.40\pm 0.07$ & $2.72\pm 0.07$ & $0.01\pm 0.01$\\
Multijet       & $<0.05$     & $0.07\pm 0.07$ & $<0.05$\\
\hline
Background sum & $2.30\pm 0.21$ & $3.81\pm 0.17$ & $1.95\pm 0.41$\\
\hline
Data           & 6              & 15             & 4\\
\end{tabular}
\end{ruledtabular}
\end{table}

Systematic uncertainties that affect the $WW$ production cross section 
measurement are listed in Table~\ref{tab:syst}. 
In these estimates, parameters are varied within $\pm
1\sigma$ of the respective theoretical or experimental errors. 
Sources such as the trigger efficiency, electron  and muon identification 
(ID) efficiencies, jet energy scale (JES),  electron and 
muon momentum resolution, branching fraction $B(W\to \ell\nu)$, cross
section calculation of \zg\ and \ttbar\ events, and the determination 
of $W$+jet/$\gamma$ background contribute to the systematic uncertainty. 
The {\sc pythia} Monte Carlo tends to underestimate jet multiplicities, since
a parton-shower approach is used for initial and final state radiation
instead of the full matrix element. To compensate for this underestimation, 
events are re-weighted in the MC to reproduce the jet multiplicities seen in 
the 
data. The systematic
uncertainty for this approach is determined from a measurement of the $WW$
production cross section with and without the re-weighting.
The total systematic uncertainties are given in Table~\ref{tab:syst}.
The uncertainty on the luminosity measurement is 6.5\%.

\begin{table}[t]
\caption{\label{tab:syst} Systematic uncertainties for the
  $ee$, $e\mu$, and $\mu\mu$ channels.} 
\begin{ruledtabular}
\begin{tabular}{lcccccc}
 Source       & \multicolumn{6}{c}{Change in the $WW$ cross section (\%)} \\
              & \multicolumn{2}{c}{$ee$} & \multicolumn{2}{c}{$e\mu$} & \multicolumn{2}{c}{$\mu\mu$} \\
\hline
 Trigger, ID    & $+4.7$ & $-4.6$ & $+3.9$ & $-3.8$ & $+6.2$ & $-5.8$ \\
 JES            & $+3.2$ & $-3.2$ & $+1.6$ & $-1.2$ & $+7.2$  & $-4.8$ \\
 $\mu$ resolution               & $-$        & $-$    & $+4.7$ & $-2.2$ & $+10.0$ & $-4.1$ \\
 $e$ resolution           & $+4.6$ & $-2.9$ & $+1.3$ & $-1.1$ & $-$       & $-$  \\ 
 $B$($W\rightarrow \ell \nu$) & $+4.4$ & $-3.9$ & $+5.3$ & $-4.6$ & $+4.3$  & $-4.1$ \\
 $\sigma(\zg , \ttbar)$      & $+0.9$ & $-0.7$ & $+0.4$ & $-0.4$ & $+3.2$  & $-3.2$ \\
 $W$+jet/$\gamma$                    & $+4.0$ & $-4.0$ & $+3.0$ & $-3.0$ & $-$       &  $-$ \\ 
 Re-weighting            & $+4.3$ & $-4.4$ &   $-$    & $-$      & $+1.5$  & $-1.5$ \\ 
\hline
 Total                       & $+10.3$ & $-9.5$ & $+8.9$ & $-7.3$ & $+14.9$ & $-10.1$ \\    
\end{tabular}
\end{ruledtabular}
\end{table}

The cross section for $W$ boson pair production is estimated using a likelihood
method \cite{feldman, roe} with Poisson statistics. 
The cross section for each channel $\sigma_{p\bar{p}\to  W^+W^-}$ is given by
\begin{equation}
\sigma_{p\bar{p}\to W^+W^-} = \frac{N_{\rm obs} - N_{bg}}{\lint\cdot B\cdot\epsilon}\ ,
\end{equation}
where $N_{\rm obs}$ is the number of observed events, $N_{bg}$ is the expected
background, \lint\ is the integrated luminosity, $B$ is the
branching fraction for $W\rightarrow \ell\nu$, and $\epsilon$ is the efficiency
for the signal. The
likelihood for $N_{\rm obs}$ events in the data is given by
\begin{equation}
L(\sigma_{p\bar{p}\to W^+W^-},N_{\rm obs},N_{bg},\lint,B,\epsilon) =
\frac{N^{N_{\rm obs}}}{N_{\rm obs}!}e^{-N}\ , 
\end{equation}
where $N$ is the number of signal and background events:
\begin{equation}
N = \sigma_{p\bar{p}\to W^+W^-}\cdot B\cdot\lint\cdot\epsilon + N_{bg}\ .
\end{equation}
The cross section $\sigma_{p\bar{p}\to W^+W^-}$ is estimated by minimizing
$-2\ln L(\sigma_{p\bar{p}\to W^+W^-},N_{\rm obs},N_{bg},\lint,B,\epsilon)$. 
To combine the channels, the individual likelihood functions are multiplied.
As a final result, the combined  cross section for $WW$ production at a 
center-of-mass energy of $\sqrt{s}$ = 1.96~TeV is
\begin{equation}
\sigma_{p\bar{p}\rightarrow W^+W^-} =
13.8^{+4.3}_{-3.8}(\rm stat)^{~+1.2}_{~-0.9}(\rm syst)  \pm 0.9(\rm lum)\ {\rm pb}.
\end{equation}
This value is in good agreement with the NLO calculation prediction of 
$12.0$--$13.5$ pb at $\sqrt{s}=1.96$ TeV~\cite{ellis}.

The significance for the signal observation can be estimated using the
likelihood ratio method \cite{junk}. The confidence levels for a background
only hypothesis, $CL_{B}$, is obtained using the background
expectation and the number of events observed as input.
The signal significance is extracted from $1-CL_{B}$. The probability of an 
upward fluctuation of the background is $2.3\times 10^{-7}$, which corresponds 
to 5.2 standard deviations for a Gaussian probability distribution. 

To conclude, we have measured the $W$ boson pair production cross section
in \ppbar\ collisions at $\sqrt{s}$ = 1.96~TeV. 
We observe 25 events
in the data, corresponding to integrated luminosities of 
224--252~\ipb\ depending on the final state, 
with a background expectation from non-$WW$ processes of 
$8.1\pm0.6(\rm stat) \pm 0.6 (\rm syst) \pm 0.5(\rm lum) $ events. 
The expectation for SM pair production of $W$ bosons in our data sample
is $16.6\pm0.1(\rm stat) \pm 0.6 (\rm syst) \pm 1.1(\rm lum) $ events. 
We obtain a production cross section of $\sigma_{p\bar{p}\rightarrow W^+W^-} =
13.8^{+4.3}_{-3.8}(\rm stat)^{~+1.2}_{~-0.9}(\rm syst)  \pm 0.9(\rm lum)\ {\rm
  pb}$, consistent with the NLO prediction.
The probability that the observed events are caused by a fluctuation of the
background is $2.3\times 10^{-7}$, corresponding to 5.2 standard deviations.

%
We thank the staffs at Fermilab and collaborating institutions, 
and acknowledge support from the 
Department of Energy and National Science Foundation (USA),  
Commissariat  \` a l'Energie Atomique and 
CNRS/Institut National de Physique Nucl\'eaire et 
de Physique des Particules (France), 
Ministry of Education and Science, Agency for Atomic 
   Energy and RF President Grants Program (Russia),
CAPES, CNPq, FAPERJ, FAPESP and FUNDUNESP (Brazil),
Departments of Atomic Energy and Science and Technology (India),
Colciencias (Colombia),
CONACyT (Mexico),
KRF (Korea),
CONICET and UBACyT (Argentina),
The Foundation for Fundamental Research on Matter (The Netherlands),
PPARC (United Kingdom),
Ministry of Education (Czech Republic),
Natural Sciences and Engineering Research Council and 
WestGrid Project (Canada),
BMBF and DFG (Germany),
A.P.~Sloan Foundation,
Research Corporation,
Texas Advanced Research Program,
and the Alexander von Humboldt Foundation.
%

\end{document}